\documentclass[twocolumn]{aastex631}

\usepackage{graphicx}	% Including figure files
\usepackage{amsmath}	% Advanced maths commands
\usepackage{amssymb}	% Extra maths symbols
\usepackage{color}
\usepackage{booktabs}
\usepackage{pifont}% http://ctan.org/pkg/pifont
\usepackage{soul}
\usepackage{multibib}
\newcites{Appendix}{Appendix References}
\usepackage{natbib}
\usepackage[flushleft]{threeparttablex}
\usepackage{mathrsfs}
\usepackage{tikz}
\usepackage{esint}
\usepackage{lipsum}
\usepackage{longtable}
\usepackage{pifont}% http://ctan.org/pkg/pifont
\usepackage{savesym}
\savesymbol{tablenum}
\usepackage{siunitx}
\restoresymbol{SIX}{tablenum}
\usepackage{placeins}
\usepackage{bm}

\newcommand{\appropto}{\mathrel{\vcenter{
  \offinterlineskip\halign{\hfil$##$\cr
    \propto\cr\noalign{\kern2pt}\sim\cr\noalign{\kern-2pt}}}}}

\newcommand{\Exp}[2]{\left\langle{#1}\right\rangle_{#2}}

\renewcommand{\d}[1]{\ensuremath{\operatorname{d}\!{#1}}}

\DeclareMathOperator\D{\mathcal{D}}
\DeclareMathOperator\V{\mathcal{V}}
\newcommand{\M}{\mathcal{M}}

\DeclareMathOperator\s{\text{s}}
\DeclareMathOperator\f{\text{f}}

\DeclareMathAlphabet\mathbfcal{OMS}{cmsy}{b}{n}

\renewcommand{\v}{\mathit{v}}
\renewcommand{\b}{\mathit{b}}

\renewcommand{\u}{\bm{u}}

\renewcommand{\k}{\bm{k}}
\newcommand{\x}{\bm{x}}
\newcommand{\K}{\bm{K}}
\renewcommand{\P}{\mathcal{P}_{u}}

\newcommand{\T}{\mathcal{T}}
\newcommand{\bnab}{\bm{\nabla}}

\newcommand{\St}{{\rm{St}}}
\newcommand{\C}{{\mathcal{C}}}
\renewcommand{\ol}[1]{\overline{#1}}

\journalinfo{}

\begin{document}

\correspondingauthor{\\
$^{\dagger}$James R. Beattie: \href{mailto:jbeattie@cita.utoronto.ca}{jbeattie@cita.utoronto.ca}\\
$^{\ddagger}$Isabelle Connor: \href{mailto:iconnor@ucsc.edu}{iconnor@ucsc.edu}\\
$^\star$ these authors made equal contributions and should be regarded as joint first authors for the publication.}

\title{What is the Strouhal number of turbulence driven by supernovae?}

\author[0000-0001-9199-7771]{James R. Beattie$^{\dagger,\star}$}
\affiliation{Canadian Institute for Theoretical Astrophysics, University of Toronto, 60 St. George Street, Toronto, ON M5S 3H8, Canada}
\affiliation{Department of Astrophysical Sciences, Nassau Street, Princeton University, Princeton, NJ 08544, USA}

\author[0009-0008-3986-657X]{Isabelle Connor$^{\ddagger,\star}$}
    \affiliation{Department of Astronomy and Astrophysics, University of California, Santa Cruz, 1156 High Street, Santa Cruz, CA 96054}

\author[0000-0003-2558-3102]{Enrico Ramirez-Ruiz}
    \affiliation{Department of Astronomy and Astrophysics, University of California, Santa Cruz, 1156 High Street, Santa Cruz, CA 96054}

\begin{abstract}
    The Strouhal number, $\St=t_{\rm cor}/t_{\rm out}$, measures the temporal coherence of turbulent driving relative to the outer-scale eddy turnover time. In turbulence-box models one commonly sets $\St=1$, although recent work by \citet{Grete2025_density_distribution} and \citet{Scannapieco2025_density_distribution} has shown that turbulence statistics, especially the mass-density distribution in compressively driven turbulence, are sensitive to this choice. In this Letter, we compute $\St$ directly from the measured two-time correlation tensor and outer-scale eddy time in stratified multiphase ISM simulations of Milky Way-like and starburst disks. We find isotropic median values $\St=0.26^{+0.30}_{-0.16}$ for the Milky Way-like model and $\St=0.25^{+0.11}_{-0.12}$ for the starburst model. These values are consistent with the picture that supernova remnants (SNRs) drive turbulence locally near $R_{\rm cool}$, where the unstable contact discontinuity in the expanding SNR sets comparable forcing and eddy times, $\St(R_{\rm cool})\approx 1$. The reconstructed scale-dependent curves reach $\St=1$ at a nearly universal outer-scale fraction, $\ell_\ast/\ell_{\rm out}\approx0.12\text{--}0.13$ ($\ell_\ast\approx25\text{--}32\,\rm{pc}$), so the standard $\St=1$ prescription is not an outer-scale model of SN-driven ISM turbulence, but a local-scale approximation tied to injection near the cooling radius of the SNR.
\end{abstract}


\section{Introduction} \label{sec:intro}

    In standard momentum-forced turbulence models, as commonly employed in idealized box simulations of astrophysical turbulence, the momentum equation is written
    \begin{align}
        \frac{\partial \rho \u}{\partial t} + \bnab\cdot\mathbb{F}_{\rho\u} = \rho\bm{f}_{\rm turb},
    \end{align}
    where \(\rho\) is the mass density, \(\u\) is the fluid velocity, \(\mathbb{F}_{\rho\u}\) is the conserved momentum-flux tensor, and $\bm{f}_{\rm turb}$ is the stochastic driving acceleration, usually concentrated at long wavelengths to emulate outer-scale forcing. A standard choice is to construct $\bm{f}_{\rm turb}$ as an Ornstein--Uhlenbeck process,
    \begin{align}
        \d{\hat{\bm{f}}_{\rm turb}(\k)} = f_0(\k)\mathcal{P}_{ij}(\zeta)\d W_i(t) - \hat{f}_{i,\rm turb}(\k)\frac{\d t}{t_{\rm cor}},
    \end{align}
    where hats denote Fourier-space quantities at wavevector \(\k\), $f_0(\k)$ sets the forcing amplitude and spectral envelope, $\mathcal{P}_{ij}(\zeta)$ fixes the solenoidal-compressive mixture through the parameter \(\zeta\), $\d W_i(t)$ is a Wiener increment, and $t_{\rm cor}$ is the forcing correlation time \citep[e.g.,][]{Eswaran1988_forcing_numerical_scheme,Schmidt2008,Schmidt2009,Federrath2008,Federrath2010_solendoidal_versus_compressive,Price2010_grid_versus_SPH,Federrath2022_turbulence_driving_module}. Writing $\bm{f}_{\rm turb} = u_{\rm out}t_{\rm out}^{-1}\ol{\bm{f}}_{\rm turb}$, $f_0(\k)=u_{\rm out}t_{\rm out}^{-3/2}\ol{f_0}(\k)$, $\d W_i(t)=t_{\rm out}^{1/2}\ol{\d W_i(t)}$, and $\d t=t_{\rm out}\ol{\d t}$ gives
    \begin{align}
        \d{\ol{\bm{f}}_{\rm turb}(\k)} = \ol{f_0}(\k)\mathcal{P}_{ij}(\zeta)\ol{\d W_i(t)} - \ol{f}_{i,\rm turb}(\k)\frac{\ol{\d t}}{\St},
    \end{align}
    where $t_{\rm out}$ is the outer-scale eddy turnover time, with characteristic velocity $u_{\rm out}$, and
    \begin{align}
        \St = \frac{t_{\rm cor}}{t_{\rm out}}
    \end{align}
    is the Strouhal number. Classically, $\St$ is defined as \(fL/U\), measuring the ratio between oscillatory and advective times in unsteady flows such as vortex shedding, where $f$ is the oscillation frequency, and $U$ and $L$ are the characteristic velocity and length scales, respectively \citep{Strouhal1878,Roshko1954,Williamson1996}. In turbulence and dynamo theory the same dimensionless idea is generalized to a correlation time divided by a turnover time, where it measures the finite memory of the flow and enters dynamo closure arguments \citep{KrauseRadler1980,Blackman2002_mean_field,Kapyla2005_strouhal_convection,Brandenburg2005_astro_dynamos,Brandenburg2008_nonlinear_dynamo}; finite decorrelation is also related to the time irreversibility of cascade statistics \citep{Grafke2015_time_irreversibility}. Thus $\St$ compares the temporal coherence of the forcing to the nonlinear turnover time of the outer-scale turbulence: $\St\to0$ is the white-noise limit, $\langle f_i(\x,t)f_j(\x',t')\rangle=\mathcal{C}_{ij}(\x,\x')\delta(t-t')$, whereas $\St\to\infty$ corresponds to forcing that is effectively frozen over an eddy time. In idealized turbulence-box calculations one commonly sets $\St=1$, i.e. $t_{\rm cor}=t_{\rm out}$, following the broader numerical-turbulence practice of imposing a finite-correlated large-scale acceleration field with an autocorrelation time comparable to the large-eddy turnover or crossing time.
    
    However, the choice to set $\St = 1$ is a degree of freedom in turbulence-box (TB) models and is not constrained by a detailed analytical or numerical calculation. Furthermore, $\St$ has never been calculated for more realistic, compressible interstellar (ISM), circumgalactic (CGM), or intracluster medium (ICM) plasmas, where turbulence models with Fourier forcing are used regularly \citep{Federrath2021,Fielding2022_ISM_plasmoids,Beattie2025_nature_astro,Mohapatra2020,Mohapatra2019_turbulent_heat_flux_ICM}. It remains unknown whether $\St = 1$ is realized at all, and if it is, on what scales. The sensitivity of density statistics to the solenoidal-compressive forcing mixture is well established \citep{Federrath2008,Federrath2010_solendoidal_versus_compressive,Federrath2012,Federrath2013b}, and recent stochastic-density models have further emphasized the role of temporal correlations \citep{Scannapieco2018,Scannapieco2024_density_fluctuations}. In particular, \citet{Grete2025_density_distribution} and \citet{Scannapieco2025_density_distribution} show that the forcing correlation time changes the density statistics of compressively driven turbulence. In their notation the relevant parameter is \(\tau_{\rm a}/\tau_{\rm e}\), or equivalently \(\lambda_{\rm a}=\ln(\tau_{\rm a}/\tau_{\rm e})\), where \(\tau_{\rm a}\) is the acceleration-field correlation time and \(\tau_{\rm e}\) is an eddy time. Long-lived compressive forcing produces broader, more skewed density PDFs and large low-density voids, while short-correlated forcing produces a narrower distribution. This matters because a number of theories for characterizing turbulence in the ISM, CGM, and IGM are built from the mass-density PDF \citep{Molina2012_dens_var,Nolan2015,Burkhart2021_stats_turb_review,Mohapatra2020b,Beattie2021_multishock}, as are local models of the star-formation rate and fragmentation in the cold ISM \citep{Krumholz2005,Hennebelle2008,Federrath2012,Semenov2016_critical_s,Burkhart2018,Khuller2021,Hennebelle2024_inefficient_SF,Guochao2026_sub_grid_density}. Any change in the mass-density PDF will therefore require revisions to models derived from TB calculations, as outlined in \citet{Scannapieco2025_density_distribution}.

    In this Letter, we make the first measurement of $\St$ in a supernova-driven, multiphase ISM using two stratified galactic disk cutout models, Milky Way (MW) and starburst (SB), previously analyzed in \citealt{Connor2025_cascading_from_the_winds}. We construct $\St$ directly from the measured temporal correlation time and the characteristic outer-scale eddy turnover time of the flow. We find projected median values $\St \approx 0.2\text{--}0.25$ in both galaxy models, implying that the measured correlation time is substantially shorter than the outer-scale turnover time. We then interpret these outer-scale values with a cooling-radius/contact-layer model, following the SN-driven turbulence framework developed in \citet{Beattie2025_large_scale_small_scale}, in which remnants inject turbulence most efficiently near the cooling radius, where $\St(\ell_{\rm inj})\approx 1$, while the larger eddies that populate the scale-free cascade evolve more slowly toward the outer scale. This model predicts $\St(\ell)\propto\ell^{-3/4}$. We test that prediction by constructing a scale-dependent Strouhal number directly from the measured velocity spectrum and find good agreement, with the commonly assumed $\St = 1$ regime reached at the nearly universal dimensionless scale $\ell_\ast/\ell_{\rm out}\approx0.12\text{--}0.13$, corresponding to an effective injection scale $\ell_\ast\approx 25\text{--}32\,\rm{pc}$. The standard $\St=1$ prescription is therefore not an outer-scale model of SN-driven ISM turbulence, but at most a local-scale approximation.

    This Letter is organized as follows: in \autoref{sec:sims} we describe the numerical simulations and define the temporal and spatial correlation scales used throughout the study. In \autoref{sec:st_numbers} we compute the global and scale-dependent $\St$, and derive a model for both statistics based on the SNR driving turbulence from deep within the cascade. Finally, in \autoref{sec:summary_and_conclusions} we summarize the main results of the study and list some of the key the limitations of the models.

    \begin{figure*}[htbp]
        \centering
       \includegraphics[width=\linewidth]{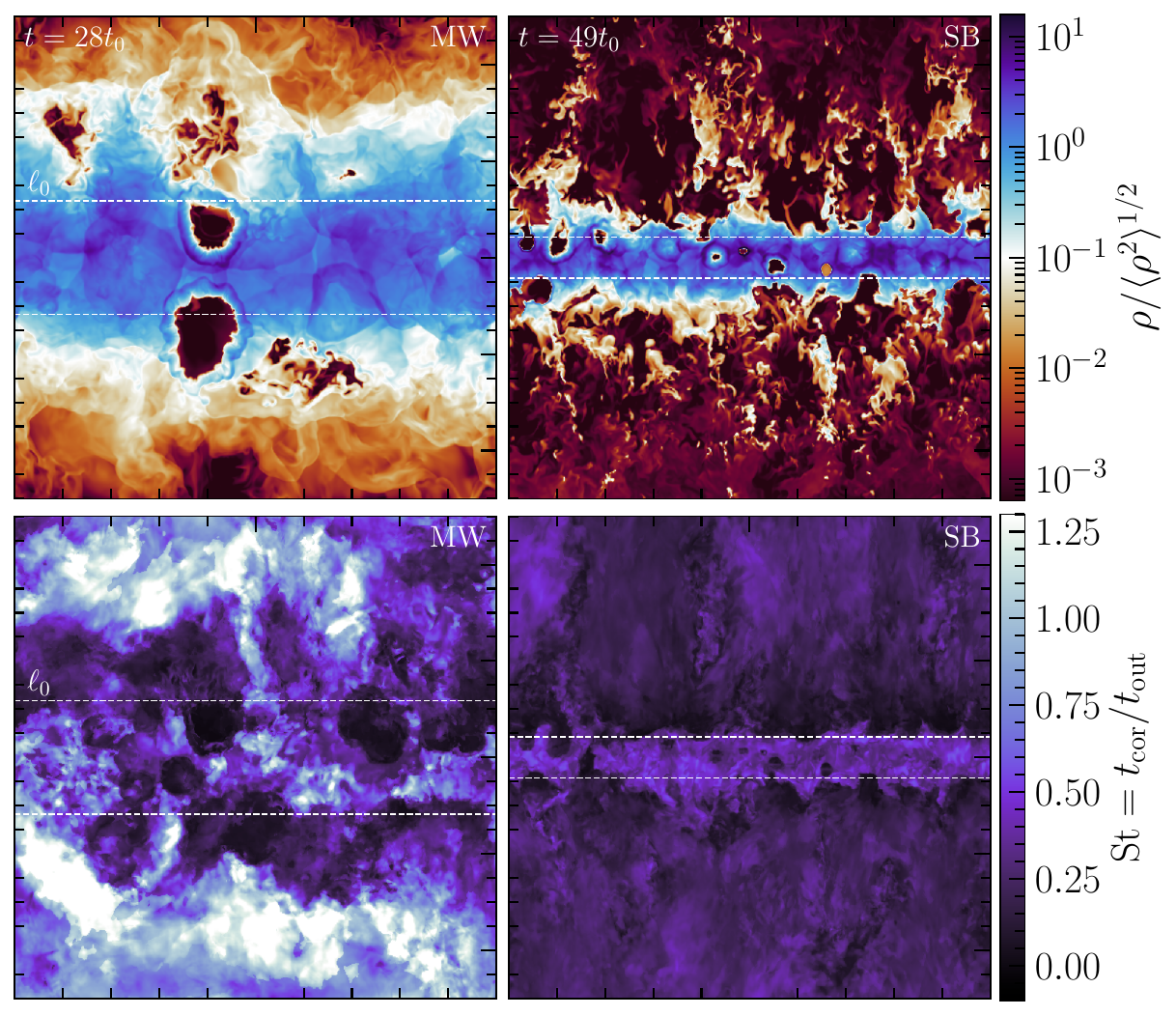}
        \caption{Two-dimensional slices parallel to $\bnab\phi$ of statistically stationary field quantities. The panels are organized by galaxy model (MW, column 1; SB, column 2), with annotations shown in the top corners of each panel. The first row is the mass density, $\rho$, and the second row is the local Strouhal number, $\St=t_{\rm cor}/t_{\rm out}$, computed over $10t_0$ of statistically steady evolution. The MW model is characterized by a thick, transonic, warm disk, and an SN explosion rate of $\gamma_{\rm SNe}/10^{-4}\,\rm{yrs} = 0.1$, whereas the SB model is characterized by a thin disk, hot, diffuse wind, and a $\gamma_{\rm SNe}$ roughly 30 times larger than the MW model.}
        \label{fig:slice}
    \end{figure*}
    
\section{Methods}\label{sec:sims}
\subsection{Gravitohydrodynamical stratified, multiphase, SN-driven turbulence model}
    For a detailed description of the numerical setup analyzed in this study, we refer the reader to \citet{Connor2025_cascading_from_the_winds} and \citet{Beattie2025_SLK41}, with additional context provided by \citet{Kolborg2022_metal_mixing_1,Kolborg2023_metal_mixing_2} and \citet{Martizzi2016}. Briefly, we use the \textsc{ramses} code \citep{Teyssier2002_ramses} to simulate ideal, stratified, gravitohydrodynamical SN-driven turbulence, incorporating a time-dependent cooling network to model the large-scale, volume-filling phases of the ISM (WNM, WIM, HIM). The setup excludes self-gravity, magnetic fields, cosmic rays, and large-scale galactic shear; a comprehensive discussion of these limitations is provided in the final section of \citealt{Beattie2025_SLK41}. This makes the simulation a controlled numerical experiment designed to isolate the impact of SN driving on stratified, multiphase turbulence. Throughout the study, we use $\u$, $\rho$, $P$, and $T$ to denote the fluid velocity, mass density, pressure, and temperature, respectively. 
    
    The domain is periodic in the plane perpendicular to the static gravitational potential gradient, $\partial_z \phi$, with outflow boundaries that allow galactic winds to escape, though these remain minimal \citep{Martizzi2016}. We use the Milky Way (MW) and starburst (SB) galaxy models and the fixed-seeding (FX) supernova seeding scheme from \citet{Connor2025_cascading_from_the_winds}. The SN explosion rate is set to $\gamma_{\rm SNe}/10^{-4}\,\rm{yrs} = 0.1$ for MW and $\gamma_{\rm SNe}/10^{-4}\,\rm{yrs} = 3.0$ for SB. The parameterization of $\phi$ and the total mass are chosen such that, for MW, the gas surface density, $\Sigma_{\rm gas} = \SI{5}{M_\odot / pc^2}$, is similar to that of the present-day solar neighborhood value (\citealp{McKee_solarneigh}; \citealp[see also][]{Martizzi2016}), and for SB, $\Sigma_{\rm gas} = \SI{50}{M_\odot / pc^2}$, analogous to a galaxy on the low end of typical starbursts (see \citealp{Kennicutt_MW_SB}). The simulation uses a domain discretization of $512^3$ on $L^3 = 1\,(\rm{kpc})^3$, which resolves scales from $1\,\rm{kpc}$ down to $\approx 40\,\rm{pc}$ before numerical diffusion truncates the dynamics \citep{Malvadi2023_numerical_diss,Beattie2025_bulk_viscosity,Grehan2025_numerical_resistivity}. Thus the cooling radius and contact layer of each individual supernova remnant are not explicitly resolved. Instead, the simulations use the subgrid feedback model of \citet{Martizzi2015}, calibrated to inject the appropriate SN energy and momentum when the remnant evolution is unresolved, consistent with shell-formation and terminal-momentum studies \citep{Thornton1998_SN_energy,Kim2015_SN_momentum,Walch2015_SN_energy_momentum,Iffrig2015_SN_molecular_clouds,Gentry2017_clustered_SNe,Gentry2019_clustered_SNe,Karpov2020_metallicity_patterns}. In the framework of \citet{Beattie2025_large_scale_small_scale}, this means that the deposited blast waves enter the resolved flow already at, or near, the unstable radiative/contact-layer stage, where they can immediately generate turbulence on the smallest dynamically available scales.

    In \autoref{fig:slice} we show two-dimensional slices through the center of the domain, parallel to $\bnab\phi$, for the two galaxy models (MW in the left column, and SB in the right column). The first row shows the mass density, $\rho$, and the second row shows the local Strouhal number, $\St=t_{\rm cor}/t_{\rm out}$, computed over $10t_0$ of statistically steady evolution. The differences between the MW and SB models are apparent in \autoref{fig:slice}, and have been discussed in detail in \citet{Connor2025_cascading_from_the_winds} and also \citet{Kolborg2022_metal_mixing_1,Kolborg2023_metal_mixing_2}. The MW model features a thick, transonic $(\M \approx 1)$, warm disk $(T \approx 10^4\,\rm{K})$ that extends throughout most of the vertical domain. In contrast, the SB model has a thin, dense disk and hot $(T \approx 10^6\,\rm{K})$, subsonic winds $(\M < 1)$ that occupy a significant portion of the simulation domain. The SB model shows filamentary structure due to many SNe exploding at the disk scale height, ejecting dense material into the winds which cools and falls back to the disk. Overall, this dichotomy between hot, wind-dominated and warm, disk-dominated statistics between the two models makes for a robust study across two significantly different ISMs.

\subsection{Definition of correlation scales} 
\label{sec:defs}

    \begin{table*}
    \caption{Characteristic timescales and correlation times for each model, averaged over the statistically steady state.}
    \label{tab:st}
    \setlength{\tabcolsep}{8pt}
    \renewcommand{\arraystretch}{0.95}
        \begin{tabular*}{\textwidth}{@{\extracolsep{\fill}}cccc@{}}
        \hline
        \hline
        Symbol & Definition & MW & SB \\
         (1) & (2) & (3) & (4) \\
        \hline
        \hline
        \multicolumn{4}{c}{\textbf{Temporal scales (in Myr)}} \\
        \hline
        $t_0$  & eddy turnover time on the gaseous scale height & $3.9\pm0.8$ & $0.37\pm0.02$ \\
        $t_{\rm out}$ & eddy turnover time on the outer scale & $7\pm1$ & $2.4\pm0.2$ \\[0.5em]
        \multicolumn{4}{c}{\textbf{Isotropic integral correlation times} (\autoref{eq:isotropic_tcor})} \\[0.5em]
        $t_{\rm cor}$ & isotropic & $1.80^{+2.07}_{-1.11}$ & $0.61^{+0.26}_{-0.29}$ \\
        $\St$ & $t_{\rm cor}/t_{\rm out}$ & $0.26^{+0.30}_{-0.16}$ & $0.25^{+0.11}_{-0.12}$ \\[0.5em]
        \multicolumn{4}{c}{\textbf{Projected integral correlation times} (\autoref{eq:project_correlations_1})} \\[0.5em]
        $t_{\rm cor}^{\perp}$ & projected perpendicular to $\bnab\phi$ & $1.51^{+1.90}_{-0.94}$ & $0.57^{+0.31}_{-0.29}$ \\
        $\St^{\perp}$ & $t_{\rm cor}^{\perp}/t_{\rm out}$ & $0.22^{+0.27}_{-0.13}$ & $0.24^{+0.13}_{-0.12}$ \\
        $t_{\rm cor}^{\parallel}$ & projected parallel to $\bnab\phi$ & $1.72^{+3.17}_{-1.20}$ & $0.53^{+0.56}_{-0.30}$ \\
        $\St^{\parallel}$ & $t_{\rm cor}^{\parallel}/t_{\rm out}$ & $0.25^{+0.45}_{-0.17}$ & $0.22^{+0.23}_{-0.12}$ \\[0.5em]
        \multicolumn{4}{c}{\textbf{Diagonal integral time tensor components}} \\[0.5em]
        $t_{\rm cor}^{xx}$ & in-plane diagonal & $1.34^{+1.76}_{-0.88}$ & $0.49^{+0.54}_{-0.26}$ \\
        $\St^{xx}$ & $t_{\rm cor}^{xx}/t_{\rm out}$ & $0.19^{+0.25}_{-0.13}$ & $0.20^{+0.22}_{-0.11}$ \\
        $t_{\rm cor}^{yy}$ & in-plane diagonal & $1.34^{+2.06}_{-0.88}$ & $0.50^{+0.54}_{-0.27}$ \\
        $\St^{yy}$ & $t_{\rm cor}^{yy}/t_{\rm out}$ & $0.19^{+0.29}_{-0.13}$ & $0.21^{+0.23}_{-0.11}$ \\
        $t_{\rm cor}^{zz}$ & out-of-plane diagonal & $1.72^{+3.17}_{-1.20}$ & $0.53^{+0.56}_{-0.30}$ \\
        $\St^{zz}$ & $t_{\rm cor}^{zz}/t_{\rm out}$ & $0.25^{+0.45}_{-0.17}$ & $0.22^{+0.23}_{-0.12}$ \\
        \hline
        \multicolumn{4}{c}{\textbf{Spatial scales (in pc)}} \\
        \hline
        $L$ & domain size & $1000$ & $1000$ \\
        $\ell_0$ & gaseous scale height & $117.7 \pm 0.5$ & $42.4 \pm0.1$ \\
        $\ell_{\rm out}$ & outer scale of the turbulence & $198 \pm 14$ & $270 \pm 12$ \\
        \hline
        $\langle u^2 \rangle_{\V}^{1/2}\,(\rm{km}\,\rm{s}^{-1})$ & global velocity dispersion & $28.9\pm6.1$ & $109.6\pm7.0$ \\
        \hline
        \hline\\[-1.25em]
        \end{tabular*}
    \tablecomments{All reported values in this table are computed over the statistically stationary state of the SN-driven turbulence (see Section 2.3 of \citealt{Connor2025_cascading_from_the_winds}), unless otherwise stated. The temporal quantities are given in units of Myr and the spatial quantities in units of pc. The first temporal row is the eddy turnover time on the gaseous scale height, $t_0=\ell_0/\langle u^2\rangle_{\V}^{1/2}$ (computed in \citet{Connor2025_cascading_from_the_winds}), and the second is the outer-scale eddy turnover time, $t_{\rm out}$ (\autoref{eq:t_out}). Correlation times are reported as the median with $16^{\rm{th}}$--$84^{\rm{th}}$ percentile ranges across the sampled grid cells. The projected quantities are formed cell by cell, $t_{\rm cor}^{\perp}(\x)=[t_{\rm cor}^{xx}(\x)+t_{\rm cor}^{yy}(\x)]/2$ and $t_{\rm cor}^{\parallel}(\x)=t_{\rm cor}^{zz}(\x)$, and the isotropic quantity is formed as $t_{\rm cor}(\x)=[t_{\rm cor}^{xx}(\x)+t_{\rm cor}^{yy}(\x)+t_{\rm cor}^{zz}(\x)]/3$, before taking spatial percentiles. Therefore the tabulated percentiles of $t_{\rm cor}$ and $t_{\rm cor}^{\perp}$ are not generally equal to the arithmetic averages of the separately tabulated percentiles of $t_{\rm cor}^{xx}$, $t_{\rm cor}^{yy}$, and $t_{\rm cor}^{zz}$. The rows labeled $\St$ give the corresponding Strouhal numbers obtained by dividing the tabulated correlation-time percentiles by $t_{\rm out}$. The spatial outer scales are computed from the velocity power spectrum using the integral definition in \autoref{sec:defs}.}
\end{table*}

    We define the outer scale of the turbulence, which is not necessarily the same as the energy injection scale in SN-driven turbulence \citep[e.g.,][]{deAvillez2005_global_ISM,2006ApJ...653.1266J,2012ApJ...750..104H,Gatto2015_SNe_driven_ISM,2017A&A...604A..70I}, as
    \begin{align}
        \frac{\ell_{\rm out}}{L} = \frac{1}{\Exp{u^2}{\V}}\int_{0}^{\infty}\d{k}\;\frac{\mathcal{P}_{u}(k)}{k},
    \end{align}
    where \(L\) is the domain size, \(k=|\k|\) is the Fourier wavenumber, \(\mathcal{P}_{u}(k)\) is the spherically averaged velocity power spectrum of the rest-frame velocity, and \(\left\langle\cdot\right\rangle_{\V}\) denotes a volume average over the simulation domain. This gives an outer-scale eddy turnover time, 
    \begin{align}\label{eq:t_out}
        t_{\rm out} = \frac{\ell_{\rm out}}{\left\langle u^2 \right\rangle_{\V}^{1/2}}.
    \end{align}
    Before forming temporal correlations, we subtract a mean velocity,
    \begin{align*}
        \delta u_i(\bm{x},t) = u_i(\bm{x},t)-\left\langle u_i\right\rangle,
    \end{align*}
    where \(\left\langle \cdot \right\rangle\) is the mean, which may be over space, time, or both (in our case \(\left\langle \cdot \right\rangle \rightarrow \left\langle \cdot \right\rangle_{\x,t}\)). We then define the Eulerian two-time correlation tensor at fixed position,
    \begin{align}\label{eq:time_correlation_function}
        \mathcal{R}_{ij}(\bm{x},t) = \left\langle \delta u_i(\bm{x},\tau)\delta u_j(\bm{x},\tau+t) \right\rangle_{\tau},
    \end{align}
    where $\bm{x}$ is a position vector and the average is taken over the reference time $\tau$ at fixed $\bm{x}$; implementation details, including the local normalization and lag normalization used in the measurements, are given in \autoref{app:correlation_implementation}. To place the diagonal and off-diagonal components on the same footing, we define the normalized correlation tensor,
    \begin{align}
        \widetilde{\mathcal{R}}_{ij}(\bm{x},t)
        =
        \frac{\mathcal{R}_{ij}(\bm{x},t)}
        {\left\langle \delta u_i^2(\bm{x},\tau) \right\rangle_{\tau}^{1/2}
        \left\langle \delta u_j^2(\bm{x},\tau) \right\rangle_{\tau}^{1/2}},
    \end{align}
    and then the correlation-time tensor,
    \begin{align}\label{eq:correlation_time}
        t_{\rm cor}^{ij}(\bm{x}) = \int_0^{t_{ij}^{(0)}(\bm{x})}\d{t}\; \widetilde{\mathcal{R}}_{ij}(\bm{x},t),
    \end{align}
    where $t_{ij}^{(0)}(\bm{x})$ is the first zero crossing of the normalized correlation.
    This defines a tensor-valued field, $t_{\rm cor}^{ij}(\bm{x})$, at each position in the domain.
    Unlike turbulence-box calculations, where the acceleration-field correlation time is imposed directly, here \(t_{\rm cor}\) is inferred from the velocity response in the gaseous disk selected by the \(z\)-range, where the statistics are dominated by SN driving.
    We therefore interpret \(t_{\rm cor}\) as an effective forcing decorrelation time for the SN-driven flow.\footnote{\label{fn:driving} A direct driving correlation could be constructed from the baroclinic source term emphasized by \citet{Beattie2025_large_scale_small_scale}, e.g., the forcing in SN-driven turbulence is from the coupling between the vorticity and the misalignment at the unstable layer, rather than an explicit momentum-coupled $\bm{f}_{\rm turb}$.
    With \(\bm{B}\equiv\bnab P\times\bnab\rho/\rho^2\), one could measure \(C_{ij}(t)=\langle\omega_i(\tau)B_j(\tau+t)\rangle\), correlating vorticity with the baroclinic torque at the unstable contact discontinuity. We leave the forcing-correlation measurement for future work that focuses specifically on the vorticity-baroclinic co-spectrum, which we discuss further in \autoref{sec:limitations}}
    Because the flow is stratified by the background gravitational potential, it is natural to separate the correlation structure parallel and perpendicular to $\bnab\phi$. From \autoref{eq:correlation_time} we therefore construct projected correlation times along $\bnab\phi$ and across $\bnab\phi$,
    \begin{align}\label{eq:project_correlations_1}
        t_{\rm cor}^{\perp}(\bm{x}) = \frac{1}{2}\mathcal{P}_{ij}^{\perp}t_{\rm cor}^{ij}(\bm{x}),
        \qquad
        t_{\rm cor}^{\parallel}(\bm{x}) = \mathcal{P}_{ij}^{\parallel}t_{\rm cor}^{ij}(\bm{x}),
    \end{align}
    where 
    \begin{align}\label{eq:project_correlations_2}
        \hat{\bm{z}} &= \bnab\phi/|\bnab\phi|, \nonumber\\
        \mathcal{P}^{\parallel}_{ij} &= \hat{z}_i \hat{z}_j, \quad \mathcal{P}^{\perp}_{ij} = \delta_{ij} - \hat{z}_i \hat{z}_j,
    \end{align}
    are the projections onto each set of coordinates, with \(\delta_{ij}\) the Kronecker delta and repeated Cartesian indices summed. The perpendicular projection contains the two in-plane directions. We find that the in-plane diagonal components satisfy $t_{\rm cor}^{xx}\approx t_{\rm cor}^{yy}$ in all models, so in the main text we represent the perpendicular correlation with the projected in-plane average. Finally, we define the isotropic correlation scale
    \begin{align}\label{eq:isotropic_tcor}
         t_{\rm cor}(\bm{x}) = \frac{1}{3}{\rm{tr}}\left\{ t_{\rm cor}^{ij}(\bm{x}) \right\},
    \end{align}
    as the trace of the correlation time tensor. The full tensor is reported in \autoref{tab:st} and \autoref{app:extra_corrs}. Related velocity and dynamo correlation scales in SN-driven ISM turbulence have been discussed in \citet{Gent2021_supernova_turbulence_and_dynamo}, \citet{Hollins2018}, and \citet{Chamandy2020_SN_driven_turbulence}.

    \begin{figure*}[htbp]
        \centering
        \includegraphics[width=\linewidth]{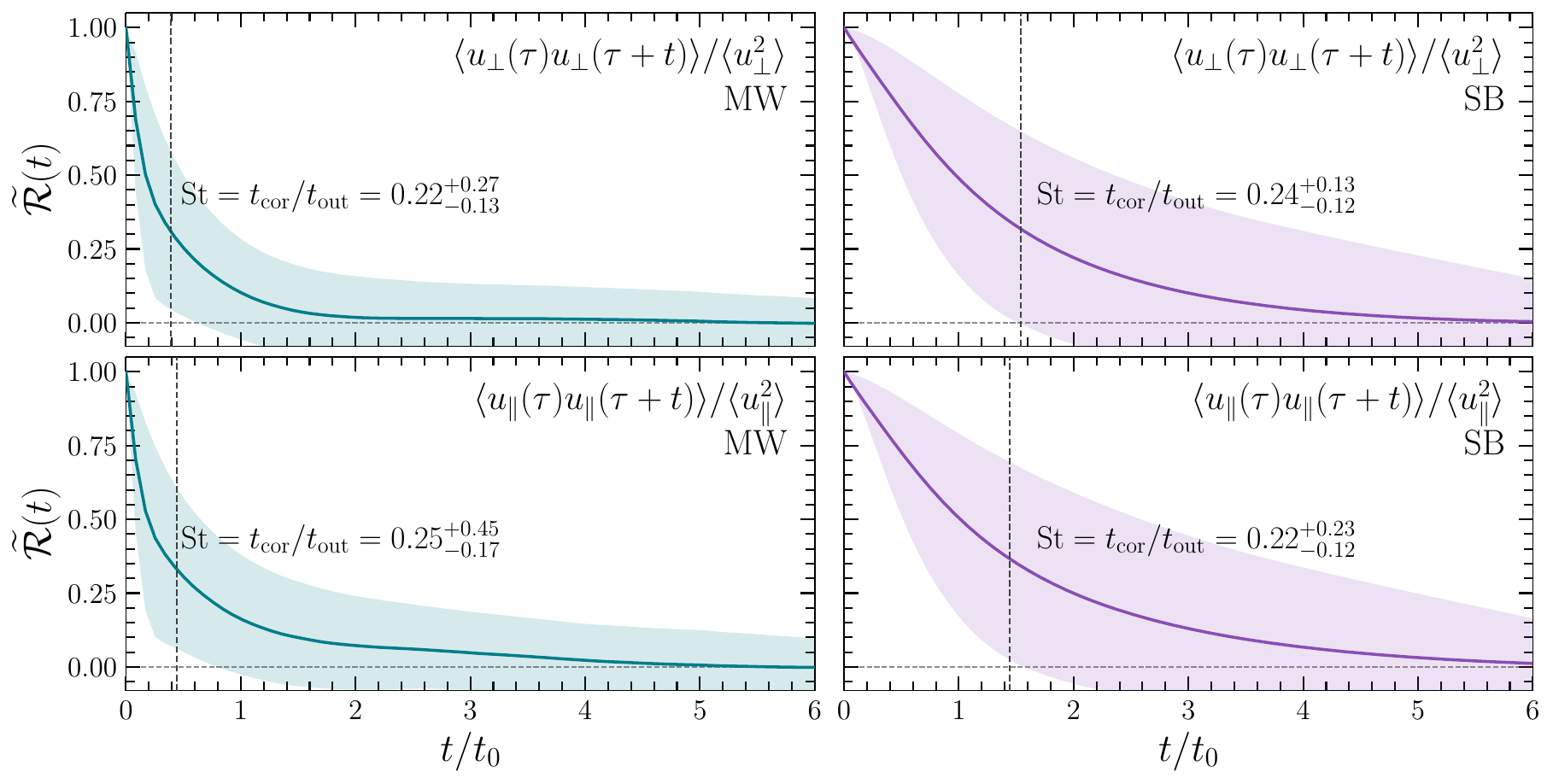}
        \caption{The normalized temporal correlation function, $\widetilde{\mathcal{R}}(t)$ (\autoref{eq:time_correlation_function}), as a function of the time lag, $t$, in units of the eddy turnover time on the gaseous scale height, $t_0$. Columns show the MW and SB galaxy models, while rows show the projected velocity components perpendicular and parallel to $\bnab\phi$. The perpendicular velocity correlation is computed from the in-plane diagonal components, $(\widetilde{\mathcal{R}}_{xx}+\widetilde{\mathcal{R}}_{yy})/2$, and the parallel velocity correlation is $\widetilde{\mathcal{R}}_{zz}$. Solid curves show the spatial mean, shaded regions show the spatial $16^{\rm{th}}$--$84^{\rm{th}}$ percentile interval, and dot-dashed vertical lines mark the median integral time, $t_{\rm cor}$. Panel annotations give the corresponding median $\St=t_{\rm cor}/t_{\rm out}$ with $16^{\rm{th}}$--$84^{\rm{th}}$ percentile ranges.}
        \label{fig:time_correlation_funcs_main}
    \end{figure*}
\section{Strouhal Numbers}\label{sec:st_numbers}
\subsection{The global Strouhal number}\label{ssec:st_glob}
    \subsubsection{Numerical results} \label{sssec:st_numerical}
    We compute \autoref{eq:time_correlation_function} and \autoref{eq:correlation_time} for the diagonal fluctuation components, $\langle \delta u_i(\tau)\delta u_i(\tau+t) \rangle$, of the correlation tensor and then form the projected parallel and perpendicular quantities defined by \autoref{eq:project_correlations_1} and \autoref{eq:project_correlations_2}. \autoref{fig:time_correlation_funcs_main} shows the projected in-plane velocity, $u_\perp$, and out-of-plane velocity, $u_\parallel$, for both galaxy models; the off-diagonal components are substantially weaker (see \autoref{tab:st} and \autoref{app:extra_corrs}). The dot-dashed gray vertical lines mark the median integral times at $t_{\rm cor}/t_0$ on the horizontal axis, and the panel annotations give the corresponding $\St=t_{\rm cor}/t_{\rm out}$. In every model, component, and direction, we find $\St<1$, consistent with broader finite-memory measurements in turbulent convection and mean-field closures \citep{Blackman2002_mean_field,Kapyla2005_strouhal_convection,Brandenburg2005_astro_dynamos,Brandenburg2008_nonlinear_dynamo} and with the expectation of \citet{Grete2025_density_distribution}.

    The projected median values are similar across the two models: $\St^{\perp}=0.22^{+0.27}_{-0.13}$ and $\St^{\parallel}=0.25^{+0.45}_{-0.17}$ for MW, and $\St^{\perp}=0.24^{+0.13}_{-0.12}$ and $\St^{\parallel}=0.22^{+0.23}_{-0.12}$ for SB. The $16^{\rm{th}}$--$84^{\rm{th}}$ percentile ranges reflect spatial variation in the local correlation-time field, but the median values are consistently sub-unity. Interpreted in the forcing notation of \citet{Grete2025_density_distribution} and \citet{Scannapieco2025_density_distribution}, these medians roughly correspond to \(\lambda_{\rm a}\sim\ln \St\approx -1.6\,\text{--} -1.4\), i.e. the short-correlation side of their parameter space, although their parameter refers to the imposed acceleration field rather than the measured velocity field. The measured outer-scale values to be explained are therefore $\St(\ell_{\rm out})<1$ with typical values near $0.2\text{--}0.25$. We now give a simple physical interpretation for these values, following the SN-driven turbulence framework of \citet{Beattie2025_large_scale_small_scale}.

    \subsubsection{\texorpdfstring{A cooling-radius/contact-layer model for SN-driven turbulence}{A cooling-radius/contact-layer model for SN-driven turbulence}} \label{sssec:st_phenom}
    \citet{Grete2025_density_distribution} studied compressively driven turbulence by varying \(\tau_{\rm a}/\tau_{\rm e}\), the analogue of $\St$ for their imposed stochastic acceleration field, and in their \S4 estimated this ratio for SN-driven turbulence by associating the forcing time with the large-scale radial expansion time of the remnant. However, earlier SN-driven ISM simulations already indicate that SN energy injection need not map onto a single effective driving scale \citep{deAvillez2005_global_ISM,2006ApJ...653.1266J,2012ApJ...750..104H,Gatto2015_SNe_driven_ISM,2017A&A...604A..70I}. Here we follow the interpretation of \citet{Beattie2025_SLK41}, \citet{Beattie2025_large_scale_small_scale}, and \citet{Connor2025_cascading_from_the_winds}, in which turbulence injection in SN-driven flows is localized instead to the unstable contact layer at the cooling radius of the remnant, $R_{\rm cool}$, where $\bnab P \times \bnab \rho/\rho^2$ generates solenoidal turbulence via the corrugated layer \citep[see also][]{Sordo2011_blastwaves_vorticity,Kapyla2018_vorticity_helicity_ISM}. In this model, the natural local condition is not $\St(\ell_{\rm out})\sim 1$, but rather $\St(\ell_{\rm inj})\sim 1$, with $\ell_{\rm inj}\sim R_{\rm cool}$. This can be understood as a consequence of instabilities at the contact discontinuity itself. We intentionally leave the specific mode ambiguous; it could involve Rayleigh--Taylor or Kelvin--Helmholtz instabilities at the contact discontinuity, although the smooth morphology of the forward shock in our simulations disfavors thin-shell modes. If the forcing correlation time is set by the inverse instability growth rate, \(\gamma_{\rm inst}^{-1}\), on $R_{\rm cool}$, then
    \begin{align}
        t_{\rm cor}\sim \gamma_{\rm inst}^{-1}\sim t_{\rm nl}(\ell_{\rm inj})\sim t_{\rm eddy}(\ell_{\rm inj}),
    \end{align}
    where \(t_{\rm nl}\) is the nonlinear turnover time of the eddy generated at the injection scale. Thus the local $\St(\ell_{\rm inj})\sim 1$ condition is not an imposed forcing prescription, but a consequence of turbulence being generated by the unstable contact discontinuity. Since the measured velocity spectrum satisfies $\P(k)\propto k^{-3/2}$, the corresponding real-space velocity scaling is $u_\ell\propto \ell^{1/4}$, and therefore
    \begin{align}\label{eq:theory_for_St}
        \St(\ell)=\frac{t_{\rm cor}}{t_{\rm eddy}(\ell)}\propto \frac{t_{\rm cor}u_\ell}{\ell}\propto \ell^{-3/4},
    \end{align}
    if $t_{\rm cor}$ is identified with the forcing correlation time.

    The corresponding outer-scale value is then
    \begin{align}
        \St(\ell_{\rm out})\sim \mathcal{C}_{\rm cool}\St(\ell_{\rm inj})\left(\frac{\ell_{\rm inj}}{\ell_{\rm out}}\right)^{3/4}
        \sim \mathcal{C}_{\rm cool}\left(\frac{R_{\rm cool}}{\ell_{\rm out}}\right)^{3/4},
    \end{align}
    where \(\mathcal{C}_{\rm cool}\) is an order-unity coefficient that absorbs how sharply one identifies the efficient injection scale with the onset of the radiative/contact-layer phase. Using the canonical range $R_{\rm cool}\approx10\text{--}30\,\rm{pc}$ \citep{Cioffi1988_SNe_cooling_radius,Blondin1998_SNe_cooling_radius,Thornton1998_SN_energy,Kim2015_SN_momentum,Martizzi2015}, together with the measured outer scales $\ell_{\rm out}\approx 198\,\rm{pc}$ for MW and $\ell_{\rm out}\approx 270\,\rm{pc}$ for SB, gives
    \begin{align}
        \St_{\rm MW}(\ell_{\rm out}) \approx 0.11\text{--}0.24, \qquad \St_{\rm SB}(\ell_{\rm out}) \approx 0.08\text{--}0.19,
    \end{align}
    for \(\mathcal{C}_{\rm cool}=1\). These estimates overlap the measured $16^{\rm{th}}$--$84^{\rm{th}}$ percentile ranges of the projected Strouhal numbers, \(\St^\perp=0.22^{+0.27}_{-0.13}\) and \(\St^\parallel=0.25^{+0.45}_{-0.17}\) for MW, and \(\St^\perp=0.24^{+0.13}_{-0.12}\) and \(\St^\parallel=0.22^{+0.23}_{-0.12}\) for SB. Thus, once the spatial variation in the measured correlation-time field is included, the cooling-radius/contact-layer model has the correct order of normalization. The empirical scale-dependent reconstruction below reaches \(\St(\ell_\ast)=1\) at \(\ell_{\ast,\rm MW}\approx25\,\rm{pc}\) and \(\ell_{\ast,\rm SB}\approx32\,\rm{pc}\). These values place the empirical injection scale near the upper end of the canonical cooling-radius range, and somewhat above it in the SB case. We therefore interpret \(R_{\rm cool,eff}\sim \ell_\ast\) as an effective radiative contact-layer scale, including order-unity uncertainty in \(\mathcal{C}_{\rm cool}\) and environmental variation, rather than as an exact measurement of a single-remnant cooling radius. The model identifies the relevant scale as the radiative cooling/contact-layer scale, not the outer scale. In this interpretation, the measured outer-scale $\St$ values are not set directly by the overall remnant lifetime, but by propagating the local condition $\St(\ell_{\rm inj})\sim 1$ from the radiative injection scale up the observed scale-free cascade.

\subsection{The Strouhal number as a function of scale} 
\label{ssec:st_ell}
\begin{figure}[t]
        \centering
        \includegraphics[width=\linewidth]{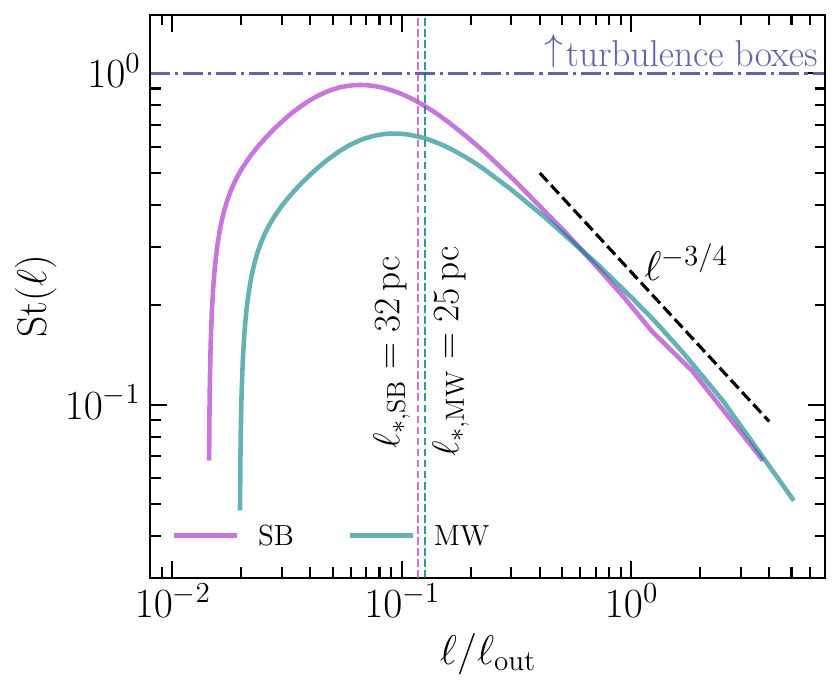}
        \caption{The Strouhal number as a function of scale, $\St(\ell)$, for the MW (teal) and SB (purple) models, derived from the measured velocity power spectrum as detailed in \autoref{ssec:st_ell}. The horizontal axis is normalized by the turbulent outer scale, $\ell_{\rm out}$, rather than by the gaseous scale height. The sloped dashed guide line shows the scale-dependent cooling-radius/contact-layer model prediction, \autoref{eq:theory_for_St}, that $\St(\ell)\propto \ell^{-3/4}$ along the scale-free cascade. The horizontal dot-dashed line marks $\St=1$, the classical turbulence-box assumption that the forcing correlation time equals the outer-scale eddy turnover time. The vertical dashed lines mark the inferred empirical crossing scales, $\ell_\ast$, where the reconstructed curves reach $\St(\ell_\ast)=1$. With this normalization the crossings occur at nearly the same dimensionless scale in both galaxy models, $\ell_\ast/\ell_{\rm out}\approx0.12\text{--}0.13$, corresponding to $\ell_{\ast,\rm MW}\approx25\,\rm{pc}$ and $\ell_{\ast,\rm SB}\approx32\,\rm{pc}$.}
        \label{fig:st_ell}
    \end{figure}

    The cooling-radius/contact-layer model above explains the measured outer-scale values by propagating a local injection-scale condition through the turbulent cascade. It therefore makes two predictions: the scale-free branch should follow $\St(\ell)\propto\ell^{-3/4}$, and the empirically inferred $\St=1$ crossing should recover an effective radiative contact-layer scale. To test these predictions, we define a scale-dependent Strouhal number by adopting the simplest assumption: $t_{\rm cor}$ is set by the driving and therefore does not vary with scale, whereas $t_{\rm eddy}$ decreases deeper into the cascade. Hence,
    \begin{align}\label{eq:st_scale}
        \St(\ell)=\frac{t_{\rm cor}}{t_{\rm eddy}(\ell)}.
    \end{align}
    We extract the turbulent velocity as a function of scale using Parseval's theorem, with \(k=1/\ell\),
    \begin{align}
        u(k) = \sqrt{\int_k^{\infty}\d{k'}\, \P(k')}
    \end{align}
    and use it to define the scale-dependent \(t_{\rm eddy}\),
    \begin{align}
        t_{\rm eddy}(k) = \frac{1}{k u(k)},
    \end{align}
    which gives \autoref{eq:st_scale}. We show the resulting $\St(\ell)$ curves in \autoref{fig:st_ell}. Along the scale-free branch, the curves agree with the predicted $\ell^{-3/4}$ scaling of \autoref{eq:theory_for_St}, as expected from the measured velocity cascade. For a pure scale-free branch normalized by an outer-scale value \(\St\), defining the crossing scale $\ell_\ast$ by $\St(\ell_\ast)=1$ gives
    \begin{align}\label{eq:st_crossing}
        \ell_\ast = \ell_{\rm out}\,\St^{4/3},
        \qquad
        \frac{\ell_\ast}{\ell_{\rm out}}=\St^{4/3},
    \end{align}
    so $\ell_\ast$ is the scale at which the extrapolated scale-free curve reaches the standard $\St=1$ regime.

    \autoref{fig:st_ell} shows that normalizing the horizontal axis by $\ell_{\rm out}$ aligns the empirical MW and SB crossing scales despite their different disk thicknesses and absolute outer scales. Along the scale-free branch, $\St(\ell)$ increases toward smaller scales following the observed $\ell^{-3/4}$ cascade, before the smallest resolved scales depart from this behavior. The vertical markers show that the reconstructed curves reach unity only near \(\ell_\ast/\ell_{\rm out}\approx0.12\text{--}0.13\), corresponding to $\ell_{\ast,\rm MW}\approx25\,\rm{pc}$ and $\ell_{\ast,\rm SB}\approx32\,\rm{pc}$. Thus the two simulations pick out nearly the same fraction of the turbulent outer scale even though the absolute crossing scales differ. This empirical reconstruction is the inverse of the cooling-radius/contact-layer prediction: starting from the measured scale-dependent $\St(\ell)$, it recovers an effective local injection scale comparable to the radiative contact-layer scale, $\ell_{\rm inj}\sim R_{\rm cool,eff}\sim \ell_\ast$, rather than the outer scale. The standard $\St = 1$ forcing prescription used in turbulence-box calculations should therefore be interpreted as a local-scale approximation, not as an outer-scale description of ISM turbulence.

\section{Summary and conclusions}\label{sec:summary_and_conclusions}
    In this Letter we make the first measurement of the Strouhal number, $\St=t_{\rm cor}/t_{\rm out}$, in a supernova (SN)-driven, multiphase ISM. The S$\St=t_{\rm cor}/t_{\rm out}$ can be interpreted as the temporal coherence of the effective forcing relative to the nonlinear turnover time of the outer-scale turbulence: it is the ratio between the forcing decorrelation time \(t_{\rm cor}\), inferred from the two-time velocity correlation tensor in \autoref{eq:time_correlation_function} and \autoref{eq:correlation_time}, and the outer-scale eddy turnover time \(t_{\rm out}\) defined in \autoref{eq:t_out}. This is analogous to the acceleration-field correlation time in standard turbulence-box models, where it is introduced as a free parameter and commonly set to unity; in contrast, here it is measured directly from the disk response to SN driving rather than prescribed. We use stratified SN-driven turbulence simulations of Milky Way (MW)- and starburst (SB)-analog galaxy models in a $1\,\rm{kpc}^3$ domain (see \autoref{sec:sims}; two-dimensional slice visualizations are shown in \autoref{fig:slice}, and the setup is detailed in \citealt{Beattie2025_SLK41} and \citealt{Connor2025_cascading_from_the_winds}). We summarize our main results as follows:
    
    \begin{itemize}
        \item In \autoref{sssec:st_numerical} we find that $\St$ in SN-driven multiphase turbulence is below unity, with projected median values $\St\approx 0.2\text{--}0.25$ in both the MW and SB models. These measurements are shown in \autoref{fig:time_correlation_funcs_main} and summarized in \autoref{tab:st}. This shows that the effective forcing in SN-driven turbulence decorrelates substantially faster than the outer-scale turbulent eddies.

        \item In \autoref{sssec:st_phenom}, we present a cooling-radius/contact-layer model based on the SN-driven turbulence framework developed in \citet{Beattie2025_SLK41}, \citet{Beattie2025_large_scale_small_scale}, and \citet{Connor2025_cascading_from_the_winds}, in which the contact discontinuity becomes unstable near the cooling radius, $R_{\rm cool}$, and the relevant condition is $\St(\ell_{\rm inj})\sim 1$ rather than $\St(\ell_{\rm out})\sim 1$. Our interpretation is that the forcing time is set by the instability growth time, $t_{\rm cor}\sim \gamma_{\rm inst}^{-1}$, while the same process sets the nonlinear time of the eddy formed at that scale, $t_{\rm nl}(\ell_{\rm inj})\sim t_{\rm eddy}(\ell_{\rm inj})$, so that $\St(\ell_{\rm inj})\sim 1$ follows directly. This leads to the scale dependence in \autoref{eq:theory_for_St} and predicts outer-scale values of order $\St(\ell_{\rm out})\sim0.1\text{--}0.3$, in good agreement with the measured $16^{\rm{th}}$--$84^{\rm{th}}$ percentile ranges in \autoref{tab:st}. SN-driven turbulence should therefore be viewed as locally forced near an effective radiative contact-layer scale, while the larger eddies that populate the cascade out to the outer scale evolve more slowly than the injection by factors of $\sim 3\text{--}10$ and extend to scales of $\ell_{\rm out}/\ell_0\approx 1.7$ in MW and $\approx 6.4$ in SB (\autoref{tab:st}).

        \item In \autoref{ssec:st_ell}, we construct a scale-dependent Strouhal number, $\St(\ell)$, using \autoref{eq:st_scale} to test the cooling-radius/contact-layer model. Along the scale-free branch, the resulting curves in \autoref{fig:st_ell} agree well with the predicted scaling $\St(\ell)\propto\ell^{-3/4}$ from \autoref{eq:theory_for_St}, consistent with the measured velocity spectrum in \citet{Beattie2025_SLK41} and \citet{Connor2025_cascading_from_the_winds}. When the curves are scaled by $\ell_{\rm out}$, the vertical markers in \autoref{fig:st_ell} show that the reconstructed $\St=1$ regime is reached at the nearly universal dimensionless scale \(\ell_\ast/\ell_{\rm out}\approx0.12\text{--}0.13\), corresponding to $\ell_{\ast,\rm MW}\approx25\,\rm{pc}$ and $\ell_{\ast,\rm SB}\approx32\,\rm{pc}$. These physical scales are best interpreted as effective radiative contact-layer scales, not exact single-remnant cooling radii. The pure scale-free conversion between \(\ell_{\rm out}\)- and \(\ell_0\)-normalized crossing scales is given in \autoref{eq:st_crossing} and \autoref{app:ell_norm}.

        \item Taken together, \autoref{fig:time_correlation_funcs_main}, \autoref{tab:st}, and \autoref{fig:st_ell} suggest that idealized turbulence-box models with $\St=1$ do not generally represent outer-scale SN-driven ISM turbulence. In the notation of \citet{Grete2025_density_distribution} and \citet{Scannapieco2025_density_distribution}, our measured outer-scale values lie on the short-correlation side of parameter space, away from the long-lived compressive-forcing cases that produce the strongest low-density voids and broadest mass-density PDFs. PDF-based models of ISM, CGM, and IGM structure, as well as star-formation rates calibrated in the long-correlation regime, may therefore need reassessment \citep{Scannapieco2018,Scannapieco2024_density_fluctuations,Scannapieco2025_density_distribution}.
        
    \end{itemize}

    More generally, the cooling-radius/contact-layer model described here should apply to turbulence driven by radiative cooling blast waves, indicating that its implications likely extend beyond the SN-driven ISM to a broader class of blast-wave-driven media.

    \subsection{Limitations}\label{sec:limitations}

    The simulations do not directly resolve the physical scales that contain the inferred $\St(\ell)\sim1$ regime: the numerical cascade reaches \(\sim40\,\rm{pc}\), while \autoref{fig:st_ell} places the empirical crossing at \(\ell_{\ast,\rm MW}\approx25\,\rm{pc}\) and \(\ell_{\ast,\rm SB}\approx32\,\rm{pc}\). Interpreting \(\ell_\ast\) as an effective radiative contact-layer scale therefore relies on extending the measured \(\P(k)\propto k^{-3/2}\) cascade below the resolved range, which should be reasonable as long as the turbulence remains SN-driven; if another driving mechanism (e.g., large-scale galactic shear or even $B \rightarrow u$ fluxes from the Lorentz force; \citealt{Grete2017}), additional microphysics, or numerical dissipation changes the slope before \(\ell_\ast\), the inferred crossing scale would shift according to \autoref{eq:st_crossing}. We also infer an effective forcing decorrelation time from the velocity response, rather than measuring the driving correlation directly. As discussed in Footnote~\ref{fn:driving}, a direct forcing signal could be constructed from the baroclinic source term emphasized by \citet{Beattie2025_large_scale_small_scale}, \citet{Sordo2011_blastwaves_vorticity}, and \citet{Kapyla2018_vorticity_helicity_ISM}. For example, with \(\bm{B}\equiv\bnab P\times\bnab\rho/\rho^2\), one could measure \(C_{ij}^{\omega B}(\x,t)=\left\langle\omega_i(\x,\tau)B_j(\x,\tau+t)\right\rangle_\tau\), which correlates vorticity with the baroclinic torque generated at the unstable contact layer. We leave this direct forcing-correlation measurement to future work that will focus on both the co-spectrum and the characteristic scales of the forcing.

\section*{Acknowledgments}
    J.~R.~B. thanks Axel Brandenburg for discussions about the Strouhal number at the KITP -- Turbulence in the Universe workshop; this was supported in part by grant NSF PHY-2309135 to the Kavli Institute for Theoretical Physics (KITP). We further thank Peng Oh, Claude-André Faucher-Giguère, Guochao (Jason) Sun, and the CITA plasma-astro group for many enlightening discussions. J.~R.~B. and I.~C. acknowledge compute allocations rrg-ripperda and rrg-essick from the Digital Research Alliance of Canada, which were used for both running and analyzing the simulations, funding from the Natural Sciences and Engineering Research Council of Canada (NSERC, funding reference number 568580), support from NSF Award 2206756, and high-performance computing resources provided by the Leibniz Rechenzentrum and the Gauss Center for Supercomputing (grants pn76gi, pr73fi, and pn76ga). This research was also made possible through funding from the Lamat Institute \citep{2025NatAs...9.1770Q} and UC Santa Cruz through the Heising-Simons Foundation and NSF grants AST-1852393, AST-1911206, AST-2150255, and AST-2206243. The authors used OpenAI Codex \citep{OpenAI2025_Codex} for assistance with manuscript editing.

    \software{We use \textsc{ramses} \citep{Teyssier2002_ramses} for all simulations. Data analysis and visualization software used in this study includes \textsc{C++} \citep{Stroustrup2013}, \textsc{mpi}, \textsc{hdf5}, \textsc{fftw}, \textsc{numpy} \citep{Oliphant2006,numpy2020}, \textsc{numba} \citep{numba:2015}, \textsc{matplotlib} \citep{Hunter2007}, \textsc{cython} \citep{Behnel2011}, \textsc{visit} \citep{Childs2012}, \textsc{scipy} \citep{Virtanen2020}, \textsc{scikit-image} \citep{vanderWalts2014}, \textsc{cmasher} \citep{Velden2020_cmasher}, \textsc{yt} \citep{yt}, \textsc{pandas} \citep{pandas}, \textsc{joblib} \citep{joblib}, and \textsc{pyfftw} \citep{2021ascl.soft09009G}.}

\appendix

\section{Correlation-function implementation}\label{app:correlation_implementation}
    This section describes the estimator used for \autoref{eq:time_correlation_function} and \autoref{eq:correlation_time}, which produce the correlation functions in \autoref{fig:time_correlation_funcs_main}, \autoref{fig:time_correlation_funcs_app_MW}, and \autoref{fig:time_correlation_funcs_app_SB}. The temporal correlation functions were computed from a time-ordered sequence of \textsc{ramses}-derived \textsc{hdf5} snapshots containing the simulation time and the three velocity components. All snapshots are ordered by their stored simulation times and restricted to the same spatial domain. For the measurements reported in \autoref{sssec:st_numerical}, the spatial domain is the model-dependent gaseous disk slab with vertical extent set by the scale height, \(\ell_0\). The selected volume is then decomposed spatially so that the correlation sums can be accumulated in parallel.

    Before forming correlations, we subtract a single space-time mean velocity for each component, computed over the selected snapshots and selected spatial domain,
    \begin{align}
        \left\langle u_i\right\rangle_{\x,t}
        =
        \frac{1}{N_{\rm cell}N_t}
        \sum_{\x}\sum_{n=0}^{N_t-1}
        u_i(\x,t_n).
    \end{align}
    Here \(N_t\) is the number of snapshots, \(N_{\rm cell}\) is the number of selected grid cells, \(t_n\) is the simulation time of snapshot \(n\), and the sum over \(\x\) runs over the selected cells.

    We then compute the temporal correlation at every nonnegative lag supported by the sampled time series. For lag index \(m\), the physical lag is the average snapshot-time separation
    \begin{align}
        \tau_m =
        \frac{1}{N_t-m}
        \sum_{n=0}^{N_t-m-1}
        \left(t_{n+m}-t_n\right).
    \end{align}
    For this lag, only the \(N_t-m\) valid, non-wrapping snapshot pairs enter the numerator. We use the linear-correlation convention returned by a zero-padded spectral estimator, for which the correlation amplitude is normalized by the total number of snapshots, \(N_t\), rather than by the number of overlapping pairs, \(N_t-m\). The corresponding unnormalized tensor is
    \begin{align}
        C_{ij}(\tau_m)
        =
        \frac{1}{N_{\rm cell}N_t}
        \sum_{\x}
        \sum_{n=0}^{N_t-m-1}
        \left[u_i(\x,t_n)-\left\langle u_i\right\rangle_{\x,t}\right]
        \left[u_j(\x,t_{n+m})-\left\langle u_j\right\rangle_{\x,t}\right].
    \end{align}
    For the correlation-time measurements used in \autoref{eq:correlation_time}, the tensor is normalized locally before spatial averaging. That is, each cell is divided by its own temporal rms, so that
    \begin{align}
        \widetilde{\mathcal{R}}_{ij}(\x,\tau_m)
        =
        \frac{1}{N_t}
        \sum_{n=0}^{N_t-m-1}
        \frac{\left[u_i(\x,t_n)-\left\langle u_i\right\rangle_{\x,t}\right]
        \left[u_j(\x,t_{n+m})-\left\langle u_j\right\rangle_{\x,t}\right]}
        {\sigma_i(\x)\sigma_j(\x)},
    \end{align}
    with
    \begin{align}
        \sigma_i^2(\x)
        =
        \frac{1}{N_t}\sum_{n=0}^{N_t-1}
        \left[u_i(\x,t_n)-\left\langle u_i\right\rangle_{\x,t}\right]^2.
    \end{align}
    Cells with zero temporal rms in either component contribute zero to the corresponding normalized component. The normalized tensor shown in the figures is the spatial average of these cell-wise normalized correlations,
    \begin{align}
        \widetilde{R}_{ij}(\tau_m)
        =
        \frac{1}{N_{\rm cell}}\sum_{\x}
        \widetilde{\mathcal{R}}_{ij}(\x,\tau_m).
    \end{align}

    The sums above are evaluated with zero-padded one-dimensional \textsc{fftw} transforms along the time axis. For each cell and velocity component, the time series is padded to at least twice its original length before transformation. Cross products of Fourier coefficients, $\widehat{u_i}^{\,*}(\x,\omega)\widehat{u_j}(\x,\omega)$, where \(\omega\) denotes the temporal Fourier frequency, are then inverse transformed to obtain the linear, non-periodic correlation sums for all nonnegative lags. The zero padding prevents wrap-around correlations, while the spatial decomposition allows the cell-wise sums and percentile summaries to be accumulated over the full selected volume.

    The spatial distribution of integral correlation times is computed from the same cell-wise correlations. For each sampled cell, \(\widetilde{\mathcal{R}}_{ij}(\x,\tau)\) is integrated with the trapezoidal rule until the first zero crossing, with linear interpolation over the final interval. The $16^{\rm{th}}$, $50^{\rm{th}}$, and $84^{\rm{th}}$ percentiles reported in \autoref{tab:st} are then recorded over cells for \(t_{\rm cor}^{ij}\), along with the projected quantities \(t_{\rm cor}^{\perp}(\x)=[t_{\rm cor}^{xx}(\x)+t_{\rm cor}^{yy}(\x)]/2\) and \(t_{\rm cor}^{\parallel}(\x)=t_{\rm cor}^{zz}(\x)\), as defined in \autoref{eq:project_correlations_1}. The projection is applied before taking spatial percentiles, so the percentile summaries of \(t_{\rm cor}^{\perp}\) are not expected to equal the arithmetic averages of the percentile summaries of \(t_{\rm cor}^{xx}\) and \(t_{\rm cor}^{yy}\).

\section{The full temporal correlation tensor} \label{app:extra_corrs}

    \begin{figure*}
        \centering
        \includegraphics[width=\linewidth]{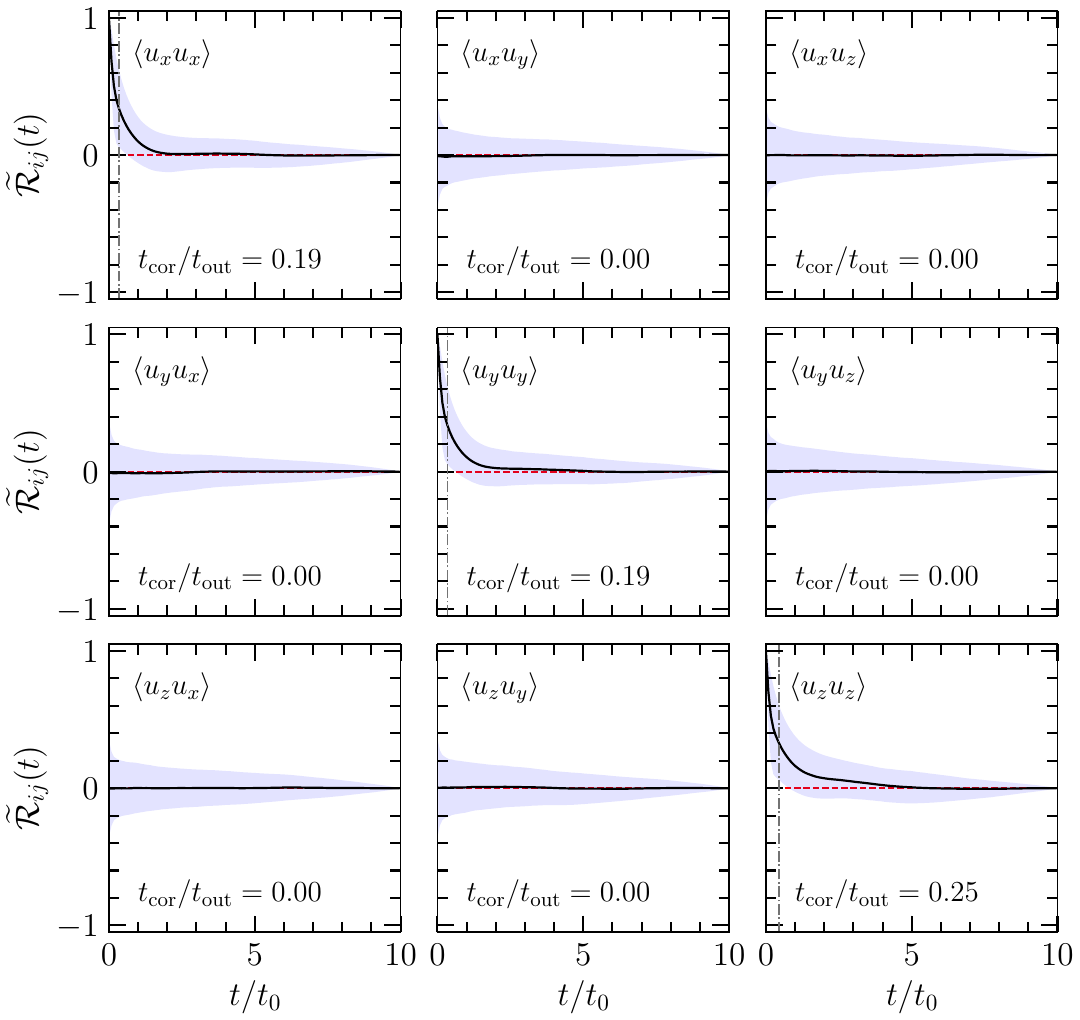}
        \caption
        {The same as \autoref{fig:time_correlation_funcs_main}, but for the full temporal correlation tensor in the MW model. Rows and columns correspond to the velocity components \(i\) and \(j\) in \(\widetilde{\mathcal{R}}_{ij}\). The diagonal components carry the dominant correlations, with median \(\St^{xx}\approx\St^{yy}\approx0.19\) and \(\St^{zz}\approx0.25\), while the off-diagonal components have zero median integral time, indicating no robust temporal coherence.}
        \label{fig:time_correlation_funcs_app_MW}
    \end{figure*}

    \begin{figure*}[t]
        \centering
        \includegraphics[width=\linewidth]{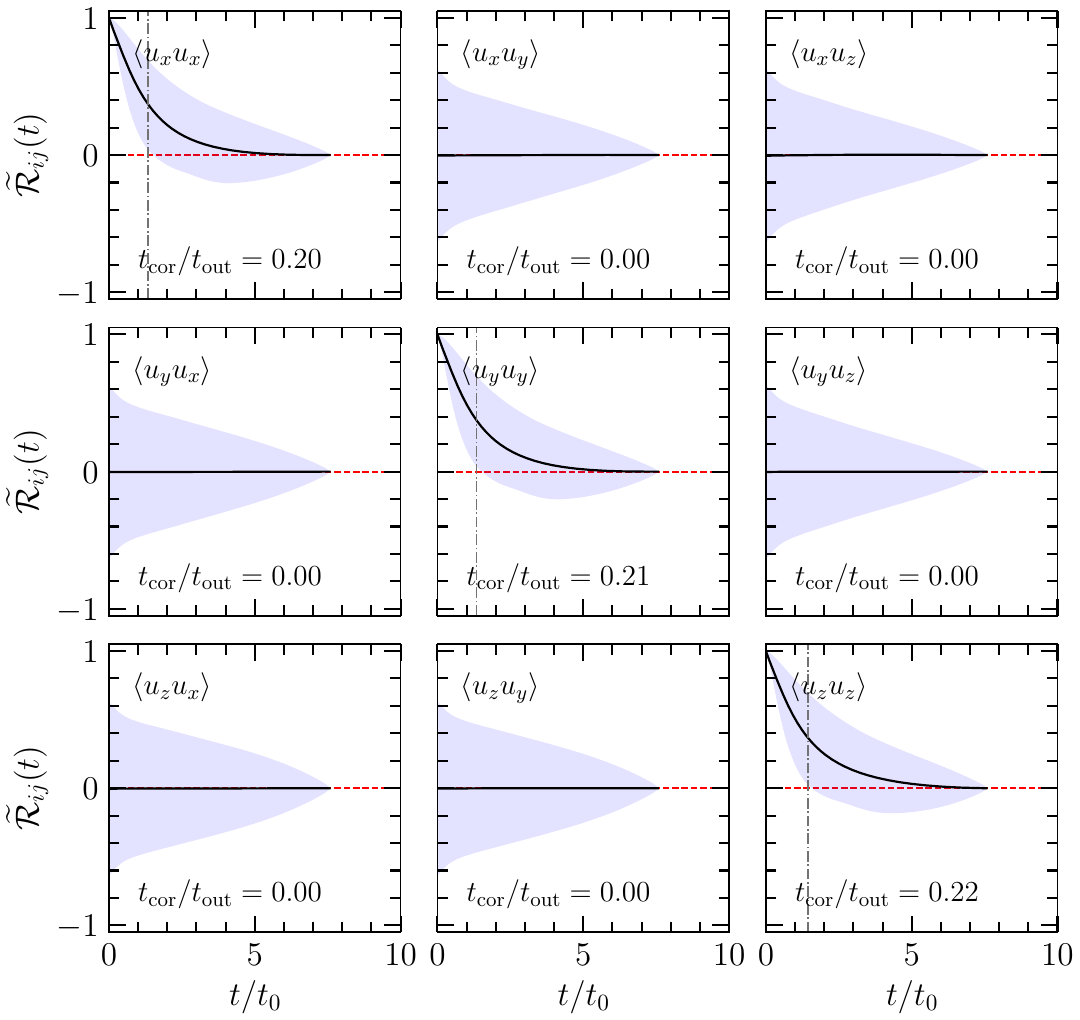}
        \caption
        {The same as \autoref{fig:time_correlation_funcs_main}, but for the full temporal correlation tensor in the SB model. Rows and columns correspond to the velocity components \(i\) and \(j\) in \(\widetilde{\mathcal{R}}_{ij}\). The diagonal components again dominate, with median \(\St^{xx}\approx0.20\), \(\St^{yy}\approx0.21\), and \(\St^{zz}\approx0.22\), while the off-diagonal components have zero median integral time.}
        \label{fig:time_correlation_funcs_app_SB}
    \end{figure*}

    \autoref{fig:time_correlation_funcs_app_MW} and \autoref{fig:time_correlation_funcs_app_SB} show the full normalized temporal correlation tensor, \(\widetilde{\mathcal{R}}_{ij}\), defined by \autoref{eq:time_correlation_function} and normalized as in \autoref{app:correlation_implementation}, for the MW and SB models, respectively. These figures extend \autoref{fig:time_correlation_funcs_main}, where only the projected in-plane and out-of-plane velocity correlations from \autoref{eq:project_correlations_1} were shown. The dominant structure is carried by the diagonal terms. In the MW model the diagonal medians are \(\St^{xx}\approx\St^{yy}\approx0.19\) and \(\St^{zz}\approx0.25\). In the SB model the corresponding diagonal medians are \(\St^{xx}\approx0.20\), \(\St^{yy}\approx0.21\), and \(\St^{zz}\approx0.22\). The full tensor therefore confirms that the component dependence is weak on the diagonal.
    
    The off-diagonal terms have median \(\St^{ij}=0\) in both models, indicating that the cross-component correlations do not produce a robust spatial-median integral time. Their $84^{\rm{th}}$ percentile values are nonzero (\autoref{tab:st}), reflecting local regions with finite cross-component coherence, but the spatial mean curves remain much weaker than the diagonal terms. The full tensor therefore supports the interpretation in \autoref{sssec:st_numerical}: the measured global \(\St\) is controlled by the autocorrelation time of the velocity components, while cross-component correlations are subdominant.

\section{Converting between outer-scale and scale-height Strouhal normalizations}\label{app:ell_norm}
    In \autoref{ssec:st_ell} and \autoref{eq:st_scale} we define the scale-dependent Strouhal number relative to the outer scale,
    \begin{align}
        \St(\ell)=\St\left(\frac{\ell}{\ell_{\rm out}}\right)^{-3/4},
    \end{align}
    where $\St$ denotes the outer-scale normalization used for a given scale-dependent curve. The exponent $-3/4$ follows from the measured velocity power spectrum and gives the model prediction in \autoref{eq:theory_for_St}, $\P(k)\propto k^{-3/2}$, which implies $u_\ell\propto \ell^{1/4}$ and hence $\St(\ell)\propto \ell^{-3/4}$. If instead one wishes to express the same curve as a function of the scale-height-normalized coordinate $\ell/\ell_0$, then
    \begin{align}
        \frac{\ell}{\ell_{\rm out}} = \frac{\ell/\ell_0}{\ell_{\rm out}/\ell_0},
    \end{align}
    so that
    \begin{align}
        \St\left(\frac{\ell}{\ell_0}\right)
        = \St\left(\frac{\ell/\ell_0}{\ell_{\rm out}/\ell_0}\right)^{-3/4}
        = \St\left(\frac{\ell_{\rm out}}{\ell_0}\right)^{3/4}\left(\frac{\ell}{\ell_0}\right)^{-3/4}.
    \end{align}
    Thus changing from $\ell/\ell_{\rm out}$ to $\ell/\ell_0$ leaves the slope unchanged, but rescales the amplitude by the factor $(\ell_{\rm out}/\ell_0)^{3/4}$. Using the values in \autoref{tab:st}, this factor is $\approx 1.48$ for the MW model and $\approx 4.01$ for the SB model, corresponding to $\ell_{\rm out}/\ell_0\approx 1.68$ and $6.37$, respectively. A useful consequence is the scale at which the extrapolated curve reaches unity. Setting $\St(\ell_\ast)=1$ gives the crossing scale used in \autoref{eq:st_crossing},
    \begin{align}
        \ell_\ast = \ell_{\rm out}\,\St^{4/3},
        \qquad
        \frac{\ell_\ast}{\ell_{\rm out}} = \St^{4/3},
        \qquad
        \frac{\ell_\ast}{\ell_0} = \frac{\ell_{\rm out}}{\ell_0}\,\St^{4/3}.
    \end{align}
    This provides the direct conversion between the crossing scale expressed in units of $\ell_{\rm out}$ and the same crossing scale expressed in units of $\ell_0$. The empirical crossings shown in \autoref{fig:st_ell} give a nearly model-independent outer-scale fraction, $\ell_\ast/\ell_{\rm out}\approx0.12\text{--}0.13$, with $\ell_{\ast,\rm MW}\approx25\,\rm{pc}$ ($\ell_\ast/\ell_0\approx 0.21$) and $\ell_{\ast,\rm SB}\approx32\,\rm{pc}$ ($\ell_\ast/\ell_0\approx 0.75$). These crossing scales are notably comparable to the canonical cooling radii of individual supernova remnants, $R_{\rm cool}\sim 10\text{--}30\,\rm{pc}$ \citep{Cioffi1988_SNe_cooling_radius,Blondin1998_SNe_cooling_radius,Thornton1998_SN_energy,Kim2015_SN_momentum,Martizzi2015}, suggesting that the scale where the reconstructed $\St(\ell)$ reaches unity is close to the physical scale where remnants become radiative and baroclinically active \citep{Beattie2025_large_scale_small_scale}.

\clearpage
\bibliography{Mar24,james_bib}
\bibliographystyle{aasjournal}

\end{document}